# NEW METHOD OF MEASURING TCP PERFORMANCE OF IP NETWORK USING BIO-COMPUTING

Ayad Ghany Ismaeel[1] and Suha Adham Abdul-Rahman[2]

[1]Department of Information Systems Engineering- Erbil Technical College- Technical Education Foundation, Erbil-Iraq
`dr_a_gh_i@yahoo.com`

[2]Ph.D. in BIO-Computing from University of Technology, Iraq
`suhaadham@yahoo.com`

## ABSTRACT

*The measurement of performance of Internet Protocol IP network can be done by Transmission Control Protocol TCP because it guarantees send data from one end of the connection actually gets to the other end and in the same order it was send, otherwise an error is reported. There are several methods to measure the performance of TCP among these methods genetic algorithms, neural network, data mining etc, all these methods have weakness and can't reach to correct measure of TCP performance.*

*This paper proposed a new method of measuring TCP performance for real time IP network using **Bio-computing**, especially molecular calculation because it provides wisdom results and it can exploit all facilities of phylogentic analysis. Applying the new method at real time on **Bio**logical **K**urdish **M**essenger **BIOKM** model designed to measure the TCP performance in two types of protocols **F**ile **T**ransfer **P**rotocol **FTP** and **I**nternet **R**elay **C**hat **D**aemon **IRCD**. This application gives very close result of TCP performance comparing with TCP performance which obtains from **Little's law** using same model (BIOKM), i.e. the different percentage of utilization (Busy or traffic industry) and the idle time which are obtained from a new method base on Bio-computing comparing with Little's law was $\cong$ **0.13%**.*

## 1. INTRODUCTION

The TCP and the **U**ser **D**atagram **P**rotocol **UDP** are both IP transport-layer protocols, but TCP can provide a reliable point-to-point communication channel that clients and servers can use to communicate with one another. There are multipurpose focus on modifying TCP performance, e.g. there is study aims to reduce the idle time before transmission at TCP by preventing timeout occurrences [14], while the others focus on maximising the utilization so the TCP can used as performance measuring because it guarantees send data from one end of the connection actually gets to the other end and in the same order it was send. The **H**yper**t**ext **T**ransfer **P**rotocol **HTTP**, FTP, IRCD [7], and **Telnet** are all examples of applications that require TCP as reliable communication protocol [10].

DOI : 10.5121/ijdps.2012.3316                  167

There are multiple methods to measure the performance of the TCP, the important of them Genetic **A**lgorithms **Gas**, **N**eural **N**etwork **NN**, **D**ata **M**ining **DM**, the problem of all these methods which are common in use for measuring the performance of TCP on IP networks they have weakness, limitation, disadvantages, so the thinking about a new method to overcome on drawbacks of these methods is an important issue.

## 2. RELATED WORK

The methods which are used for measuring the TCP performance may have disadvantages, weaknesses, limits, etc as follow [4]:

  a. **GAs**: the main weaknesses and disadvantages of this method are:
      i. The fitness equation of all the models may be similar, so convergence is slow.
      ii. It is different from other heuristic **methods** in several ways, is the fact that they find a solution through evolution.
      iii. Evolution is inductive; in nature life does not evolve towards a good solution it evolves *away* from bad circumstances. This can cause a species to evolve into an evolutionary dead end. Likewise, GAs risk finding a suboptimal solution.
      iv. A **weak method** makes few assumptions about the problem domain; hence it usually enjoys wide solving **method**. But disadvantages of such programs (as genetic algorithms) are quite **weak** without making any assumption of a problem.

  b. **NN:** the main disadvantages in this method are:
      i. Determination of the values (**biases** and **weights**) which have.
      ii. Trying with a linear model, to anticipate a group of data that are not linearly connected between them and so they can't respond efficiently to unexpected situations [13].

  c. **DM:** the main disadvantages in this method are [11]:
      i. Construct a data model in the course of data mining (like statistical ones) takes a major part of time.
      ii. The Support Vector Clustering algorithm is that the decisions function, i.e. it is impossible to estimate how much one object is "worse" than another, except for the case when one of them is determined as an outlier, and the other is not an outlier.

To avoid the problems in the methods above in measuring TCP performance it is necessary to find new method satisfying a God wisdom results like Bio-computing, the definitions of Bio-computing start with two basic principles: "**information**" and "**control**", based on these two principles by means of interaction with five related basic science disciplines [1, 4]:

  **A. Physics**: Includes physical components (physical material) and physical factors (such as temperature).

  **B. Biology and Philosophy**: Those mean Bio-computing application.

  **C. Mathematics**: Mathematical model (linear equation, and equivalence).

    **D. Engineering**: Includes genetic engineering, software engineering (the establishment of sound engineering in order to obtain a reliable software which work efficiently on real machine with low cost), and encapsulation (the addition of control information by a protocol entity to data obtained from protocol user).

The interrelated between the information and control principles means the Bio-computing sciences interconnection as show in Figure 1.

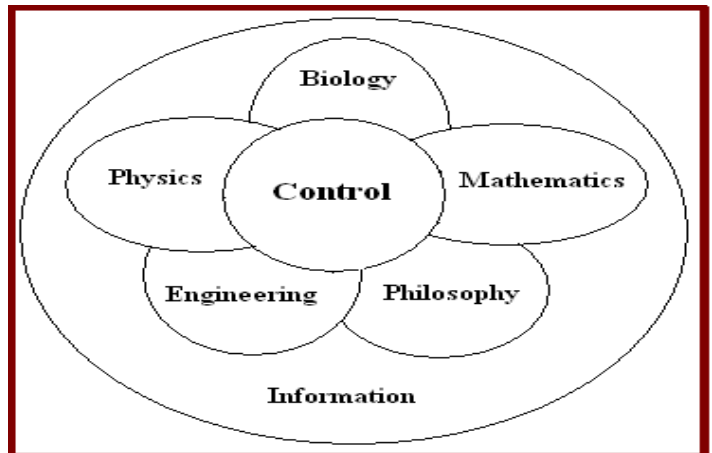

Figure 1: Bio-computing sciences interconnection [1].

Figure 1 shows the relationships between these principles (information and control) which are bases of Bio-computing, these principles give five entities as a result of the synthesis, these entities construct four disciplines (A to D above) and the basic elements of Bio-computing.

    The synthesized entities which can constitute Bio-computing can be stated below [1]:
    **i.** Bioinformatics: Is the field of science in which biology, computer science, and information technology merge to form a single discipline bio-medical informatics.
    **ii.** Molecular computing: Advanced technology for (Computer science application, and mathematics).
    **iii.** Information biotechnology: It is the synthesis between Information Technology **IT** and **B**io-**T**echnology **BT** to create an advanced technology within a specific field of application as shown in Figure 2.
    **iv.** Bio-computing philosophy.

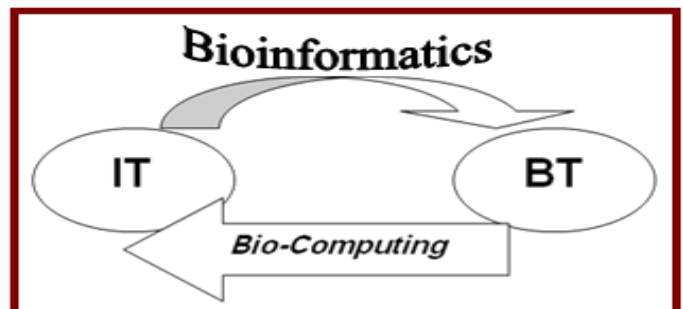

Figure 2: Shows Information Biotechnology [4].

## 3. THE MOTIVATING

TCP provides a communication channel between processes on each host system, the channel should be reliable, full-duplex, and streaming so to achieve this functionality the TCP drivers break up the session data stream into discrete segments, and attach a TCP header to each segment. An IP header is attached to this TCP packet and the composite packet is then passed to the network for delivery, this TCP header has numerous fields that are used to support the intended TCP functionality.

This research will focus on solving the measuring of TCP performance using Bio-computing especially the molecular calculation which uses to evaluate the TCP performance, because it provides wisdom results and can exploit the facilities of phylogenetic analysis (tree) for this task.

Furthermore the research goals running the proposed model efficiently by exploiting:
- The windows server (as friendly operating system) and **J**ava **D**atabase **C**onnectivity **JDBC** facilities (manipulate/control of this ultimate size of received and processed TCP packet, Sockets, protocols).
- The computer simulation of Bio-computing to obtain high TCP performance via real time network represented by phylogenetic tree (minimum tree) using phylogenetic analysis and its distance matrix strategy Neighbor Joining **NJ**.

## 4. THE ARCHITECTURE OF MEASURING TCP PERFORMANCE USING BIO-COMPUTING

The architecture of the new method which is used to measure TCP performance of IP network via Bio-computing summarize at the flowchart shown in Figure 3:

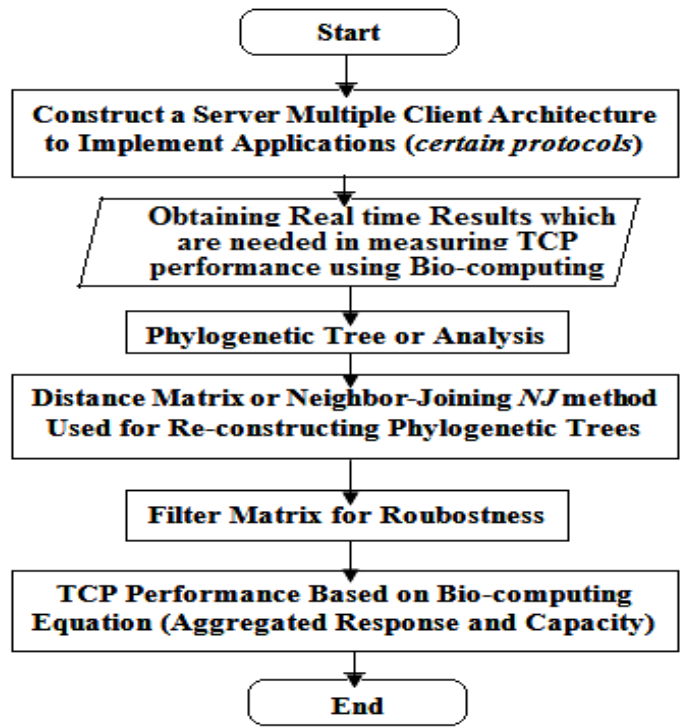

Figure 3: Flowchart of new method for measuring TCP performance using Bio-computing.

### 4.1 Server-client architecture

That means the new method will need server-client model and its applications to obtain the factors for measuring TCP performance using the new method, so the server and clients were implemented as node connected by node; it is possible to consider each path to be a tree therefore this tree is known as phylogentic tree. Neighbor joining method was used for reconstructing phylogentic tree, and then this can be taken as filter matrix which is relationship between links and paths. This filter matrix is able to select the links that are used or not. Filter matrix used for robustness which is routing algorithm property, robustness (Adaptation) with respect to failures and changing conditions (failure, congestion), then there are mathematical equations (which are explain later), like aggregate response and capacity to measure system performance.

The basic idea behind the client-server model of network communication is shown in Figure 4. The client sends messages to the server requesting service of any kind. The server responds with messages containing the desired information or takes other appropriate action. The message containing the client request is encapsulated inside a network packet and transmitted over a physical connection to the server. Conceptually, a logical connection also exists between the client and server.

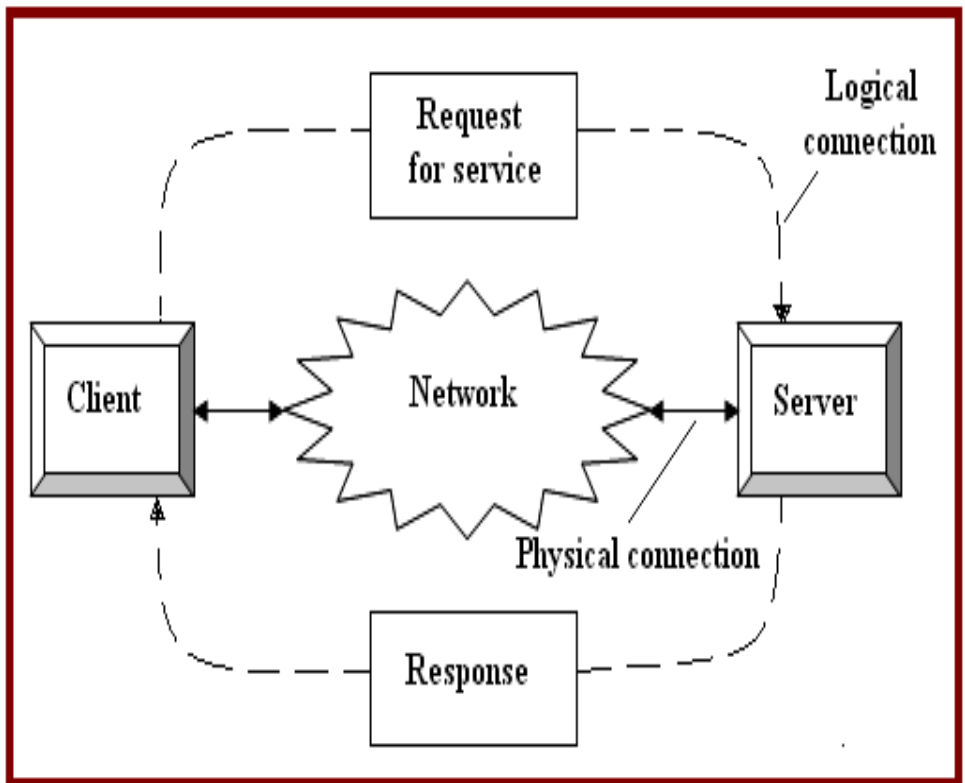

Figure 4: Server-client network architecture

The server-client messages can easily be text based and look just like the examples or they can contain data. The server must already be running for the client to communicate with it. In addition, the client must know the IP address or domain name of the server to initial communication [8]. Client-server is a useful model for sharing and managing resources on a network. File, database, print, e-mail, proxy, application, and web servers are some common examples of network applications for client-server architecture [5].

### 4.2 Phylogenetic Tree

Also called an **evolutionary tree** or **Phylogenetic analysis** is a graphical representation of the evolutionary relationship between taxonomic groups, there are many phylogenetic analysis methods as shown in Table 1. NJ method was selected for this purpose because the NJ constructs the tree by sequentially finding pairs of neighbours, which are the pairs of Operational Taxonomy Unit **OTU**s connected by a single interior node.

Table 1: Computational phylogenetic methods [1]

| Methods | Exhaustive search | Stepwise clustering |
|---|---|---|
| Character State | Maximum parsimony MP | |
| | Maximum likelihood ML | |
| Distance Matrix | Fitch-Margoliash | UPGMA |
| | | Neighbour-joining NJ |

The term phylogeny refers to the evolution or historical development of a plant or animal species, or even a human tribe or similar group. **Taxonomy** is the system of classifying plants and animals by grouping them into categories according to their similarities [3]. External (terminal) nodes are called OTUs and internal nodes are called **H**ypothetical **T**axonomic **U**nit **HTU**s. A group of taxonomy is called a cluster; as shown in Figure 5; a, the taxonomy A, B, C forms a cluster, having a common ancestor. The branching pattern that is the order of the nodes is called topology of the tree.

A rooted phylogenetic tree is a directed tree with a unique node corresponding to the most recent common ancestor of all the entities at the leaves of the tree. The most common method for rooting trees is the uses of an uncontroversial outgroup. Unrooted trees illustrate the relatedness of the leaf nodes without making assumptions about common ancestry. While un-rooted trees can always be generated from rooted one by simply omitting the root, a root cannot be inferred from an un-rooted tree without some means of identifying ancestry; this is normally done by including an outgroup in the input data or introducing additional assumptions about the relative rates of evolution on each branch. In Figure 5; b, A to E is called leaves. F to H is inferred nodes corresponding to ancestral species or molecules, branches are also called edges **[12]**.

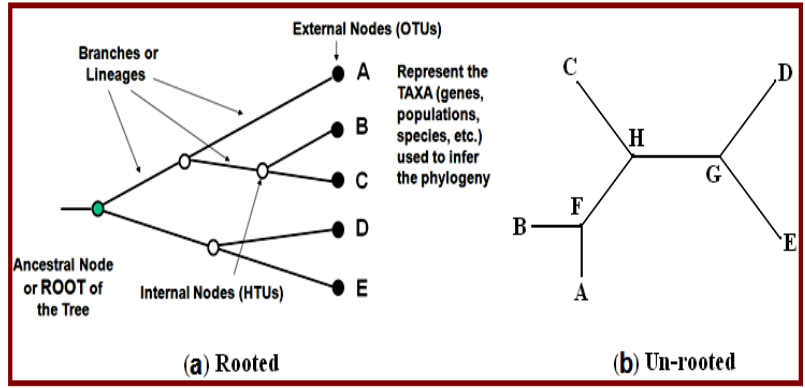

Figure 5: Structure of Rooted (a) and Un-rooted (b) phylogenetic tree.

## 4.3 Normalize the hybridized model based on Bio-computing (molecular) equation

Figure 6 clarifying the nature of relationships between Bio-computing at molecular-level technology and network routing model. The proposed model define and employ the mentioned relationships to produce an advance hybridized model, the proposed model is based on Bio-computing mathematical level, which works at molecular level, represented in phylogenetic-NJ method equations [1].

| Network Routing Model | Biological model at Molecular level (phylogenetic) |
|---|---|
| Nodes(Source, or Destination) | Organism OTU(s) |
| ↓ Synthesis | ↓ Branch |
| Nodes (sub-network) | HTU(s) |
| ↓ | ↓ Branch |
| Links | Subtree |
| ↓ | ↓ |
| Paths(exterior or interior) | Clustering |
| ↓ | ↓ |
| Clustering | Evolution Tree |
| ↓ | |
| IP Network | |

Figure 6: Analogous filling among explained models [1].

Normalize the hybridized model (mathematical + biological) based on NJ method to become more adapted to simulate the TCP performance of computer network, this simulation base on two equations:

- The total distances $U_i$, between each adjacent Organism (Operational Taxonomy Units OTUs, like *a* and *b*) can compute as shown in equation (1):

$$U_i = \sum_{\substack{b=1 \\ b \neq a}}^{N} D_{ab} \qquad \qquad \ldots \ldots \ldots \ldots \ldots \ldots (1)$$

- Will computing the total aggregate response **Q** done by equation (2):

$$Q = U_i(X_i) \text{ subject to } BS <= C \qquad \ldots \ldots \ldots \ldots (2)$$

The algorithm of Hybridized model (Normalization) shown follow [1, 4]:

---

**HYBRIDIZED MODEL (NORMALIZED) ALGORITHM**

**Input a** {a: Source node}
**Input b** {b: Destination node}
**Input D** {D: path between two network nodes each element denoted by d(a, b), row of filter matrix}
**Input L** {L: Actual links (open/close) between two nodes, each element denoted by l, column of filter matrix}
**Output** R, $R^T$, BS, C, Range, Domain, and Q

---

**R** {Filter or normalize matrix which is relationship between D and L defined as D Uses L if and only if R (d(a, b), l) = 1}
**$R^T$** {Inverse relationship for the ordered pairs resulting from the filter matrix R only exchange column with rows}
**Range** {The range for the solution which is the first position of the ordered pairs (D,L)}
**Domain** {The domain for the solution which is the second position of the ordered pairs (D, L)}
**OPL** {Output packet length which is determined in TCP/IP monitor}
**$T_T$** {Total time which is current time – server start time}
**$T_S$** {Service time which is client departure time – client start time}
**BS** {Byte size that arrive per time unit which is equal to OPL / $T_T$}
**C** {Total capacity of links which is equal to (OPL / $\sum_{C=1}^{N} TS$)}
**$D_{ab}$** {Distance between OTUs a and b}

$$U_i = \sum_{\substack{b=1 \\ b \neq a}}^{N} D_{ab} \quad \{(\text{Equation (1)}\}$$

**Total aggregate response Q = $U_i(X_i)$ subject to BS <= C** {Equation (2)}

---

## 5. EXPERIMENTAL RESULTS OF BIO-COMPUTING

Implementing the new method of measuring TCP performance using Bio-computing as shown below:

### 5.1 The requirements of implementation

The configuration of the Server-client model which has been used for BIOKM in this research can be divided into:
   A. **The Software:** The tools which are needed:

   1. **J**ava **D**evelopment Kit **JDK6:** is defined as a simple, object-oriented, distributed, interpreted, robust, secure, architecture neutral, portable, high-performance, multithreaded, and dynamic language. Therefore Java used for all research programs including client and server side programming [6, 9].
   2. **TextPad:** is a popular text editor for the Microsoft Windows family of operating systems, used as a text editor.
   3. **Microsoft Access:** is the top-notch database management system for all information management needs, from a simple address list to a complex inventory management system. It offers all the necessary tools for storing, retrieving, and interpreting the data [2].
   4. **Microsoft Excel:** is a proprietary spreadsheet application written and distributed by Microsoft. It features calculation, graphing tools, and table. Used as spreadsheet to achieve the computations of Little's law and Bio-computing algorithms.
   5. **NetBeans IDE 6.1:** is a modular, standards-based **IDE**. The NetBeans project consists of an open source IDE written in Java programming language and an application platform, which can be used as a generic framework to build any kind of application. Therefore it was used to build **J**ava **AR**chive **JAR** files used for aggregating many files into one; it is based on the ZIP file format.
   6. **Install Creator:** is a professional tool to create software installations. It offers a wizard interface that let's selects the files to include in the package, specify installation paths and then compiles the complete installation package into a compressed EXE. Therefore it was used to create **K**urd **M**essenger **S**erver **S**ide **KMSS** and **K**urd **M**essenger **C**lient **S**ide **KMCS** application installation and uninstall.
   7. **Quick Screen Capture:** is an all-in-one tool for screen capturing, image editing and organization. It can capture any part of the screen precisely in flexible ways, Used to capture the screen.

   B. **The Hardware:** Include networking (Server & Clients) components, i.e. Server-client network in star topology [4]:

   i. **The Server:** Usually called mainframe or mother computer that have a good performance and stability it used in the networking for serving other client(s). Microsoft Windows Server family which is common in use must be installed on the server computer; its installation is also like installing Microsoft Windows on client. Since the message and file exchange between users requires a centralized supervision, it will be done cross a server which controlling the transition of

    messages and maintaining most of resources for the services then joining the server with database using **J**ava **D**ata**B**ase **C**onnectivity **JDBC** for storing, retrieving, removing, and updating the data. The data mainly includes (header, user information, invited users, messages, bandwidth, number of hops, etc) in client side. Java supports database programming. JDBC is designed to provide Java programmers with a uniform way of accessing database systems. With JDBC one can access almost all current database systems such as Microsoft **S**tructured **Q**uery **L**anguage **SQL** server, Microsoft Access etc. JDBC works like this: database vendors provide drivers for their particular database system to work with JDBC driver manager. JDBC provides an abstraction layer on top of these drivers. Programs written according to JDBC API will talk to the JDBC driver manager, which in turn use the drivers that it has to talk to the actual database [6].

ii. **The Client:** Is a workstation computer or said to be PCs that add to a domain or sends the request to a server and gets the server responses. Client computer may be any type of computers that runs an operating system like Microsoft Windows supporting Microsoft Windows server. The client side and server side are communicated over socket (Socket is one end-point of a two way communication link between two programs running on the network. The java.net package provides two classes: Socket and ServerSocket that implement the client side of the connection and the server side of the connection, respectively) by receiving and transmitting messages which carries specified meanings and works as a protocol between them.

iii. **Switch (or Hub):** is an electronic device that connects multiple computers to a server via RJ45. Figure 7 shows the relationship between server-client architecture.

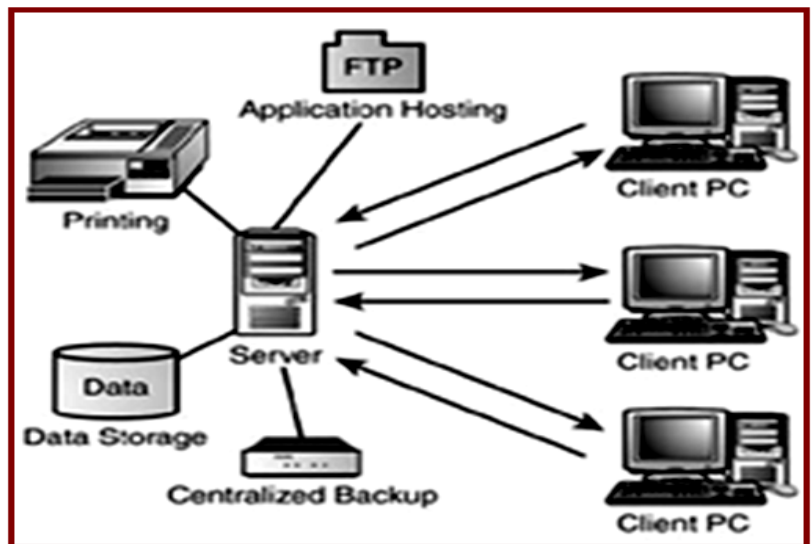

Figure 7: The relationship between server-client architecture.

### 5.2 Apply the new method of TCP performance using Bio-computing

BIOKM package design and implement in java which is used as application for server-client model, it works with Kurdish fonts to satisfy a real time IP network with two applications Kurd messenger for IRCD and Kurd messenger for FTP**,** these two protocols are connect to **K**urd **M**essenger **D**atabase **C**onnectivity **KMDBC** [4].

After installing the two main parts of BIOKM, the KMSS and KMCS**,** then implement (real time) their applications IRCD, FTP, as well as IRCD & FTP together (at the same time). That implementation gives the performance factors of BIOKM as shown in Table 2 [4]. These factors later will be use to determine TCP performance using new method (Bio-computing).

Table 2: Factors values of the application IRCD, FTP, and IRCD & FTP together

| Measurement Factors | IRCD | FTP | IRCD & FTP |
|---|---|---|---|
| No. of Online Clients | 2 | 2 | 2 |
| No. of Servers | 1 | 1 | 1 |
| Packet Sent (Packet) | 6874 | 4612 | 7341 |
| Packet Sent Length (Byte) | 6426 | 4304 | 6865 |
| Packet Received (Packet) | 6452 | 4067 | 6832 |
| Packet Receive Length (Byte) | 5868 | 3653 | 6214 |
| Total Arrival Time (Mili Second) | 32484 | 32656 | 35438 |
| Total Departure Time (Mili Second) | $1.22*10^{12}$ | $1.22*10^{12}$ | $1.22*10^{12}$ |
| Total Service Time (Mili Second) | 73328 | 111922 | 113625 |
| Total Time (Mili Second) | 111687 | 152297 | 155672 |
| Arrival Rate (Packet/Second) | 57.8 | 26.7 | 43.9 |
| Service Rate (Packet/Second) | 87.9 | 36.3 | 60.1 |
| Byte Size (Bit/Second) | 420.3 | 191.9 | 319.3 |
| Capacity (Bit/Second) | 640.2 | 261.1 | 437.5 |

### 5.3 Result evolution of Bio-computing

The result evolution of new method shown as follows [4]:

**A.** Implementing hybridized model algorithm which is (biological and mathematical) model based on NJ method as referring to in subsection 4-3, that will give the Phylogenetic tree for the BIOKM as application (server-client model), e.g. for user called (lady_engineer) as shown in Figure 8.

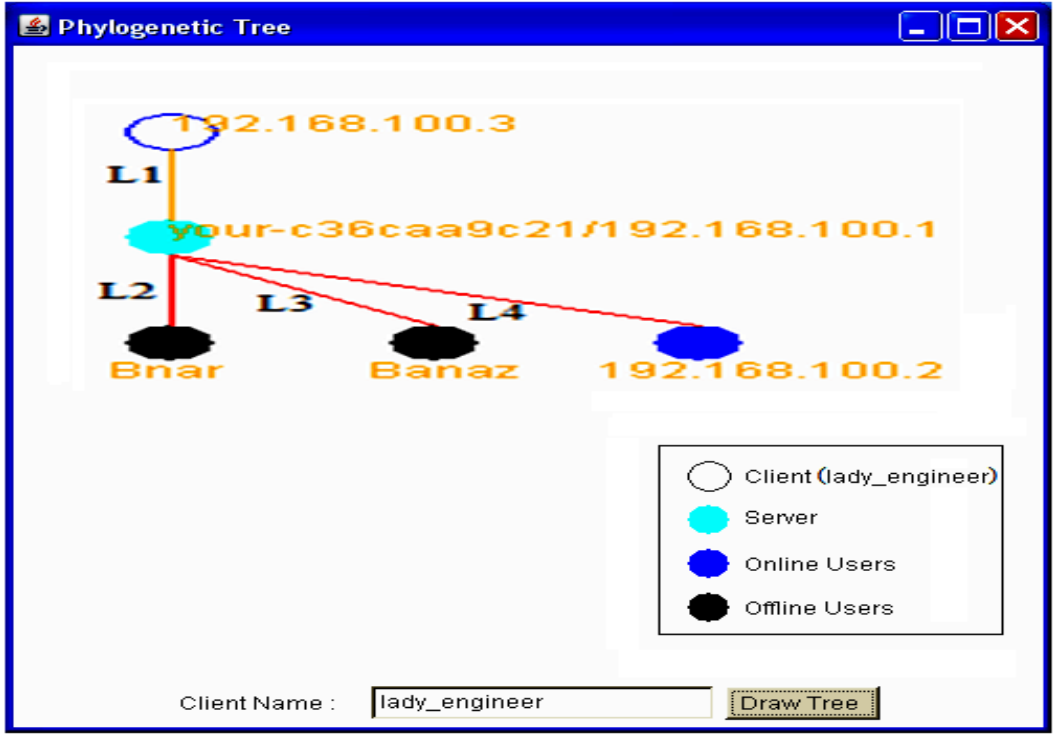

Figure 8: Phylogenetic tree for user called lady_engineer

**B.** Table 3 and Table 4 reveal the filter or normalize matrix (**R**) and transport matrix **R** (**R**$^T$) for the tree which is shown in Figure 8, that tree will use for robustness (**R**$_{li}$) where:

$$R_{li} = \begin{cases} 1 & \text{if link (l) uses source (i)} \\ 0 & \text{otherwise} \end{cases}$$

Table 3: Normalize matrix (R) for user lady_engineer

| | Matrix R | |
|---|---|---|
| | Paths | |
| | R | P 1 |
| **Links** | L 1 | 1 |
| | L 2 | 0 |
| | L 3 | 0 |
| | L4 | 1 |

Table 4: Transport matrix R (R$^T$) for user lady_engineer

| | Inverse Matrix R (R$^T$) | | | | |
|---|---|---|---|---|---|
| | Links | | | | |
| **Paths** | R$^T$ | L1 | L2 | L3 | L4 |
| | P1 | 1 | 0 | 0 | 1 |

**C.** Depending on equations (1) and (2) of hybridized model algorithm as referring to in subsection 4.3, and using the BIOKM's factors as referring to in Table 2, will obtain the results of biological applications performance (for protocols FTP, IRCD, and IRCD & FTP together) based on NJ method (Bio-computing), using Microsoft Excel as shown in Figure 9.

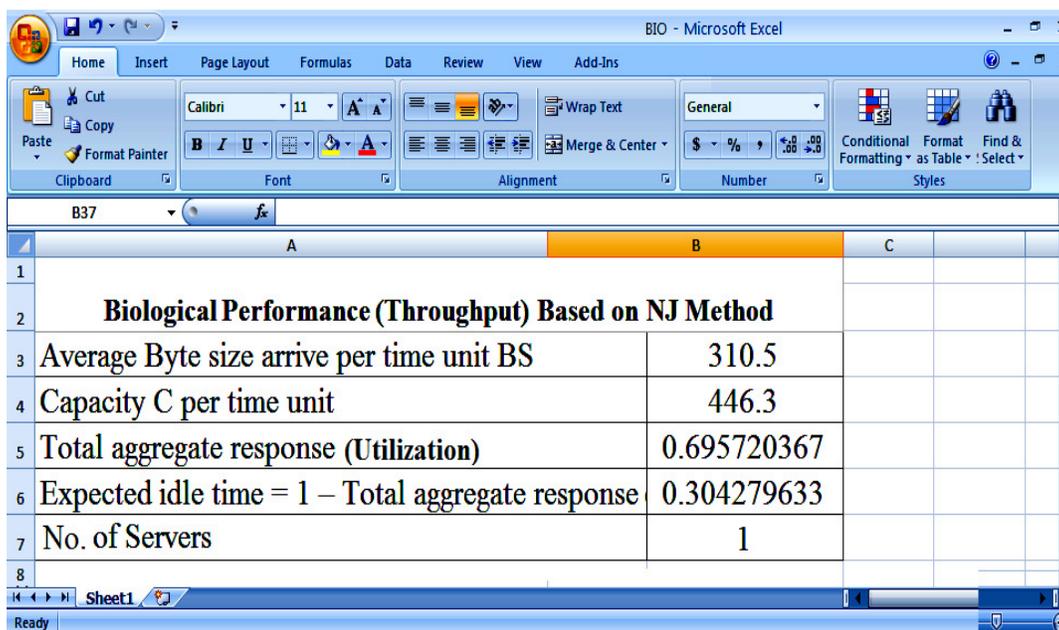

Figure 9: Shows the results of TCP performance using Biological computation (Bio-computing)

### 5.4 Discussion of the results

Little's law (Steady state system or queue theory formula) which is related to three variables, an average of units in the system at a time (**L**), the items arrive at an average rate of per unit time ($\lambda$), and average time a message spends in the system (**W**) these three variables can construct the queue formula (**L** = $\lambda$ **W**). Measuring the TCP performance uses Little's law as a single server queue (referring to as **M/M/1** model) for same Server-client model which is constructed to (Bio-computing), i.e. measuring the TCP performance by Little's law for BIOKM and its applications (FTP, IRCD, as well as FTP together with IRCD), using the factors referring to in Table 2.

Table 5 reveals comparison between the results of TCP performance measuring based on Bio-computing (new method) with Little's law for the same BIOKM's applications [4].

Table 5: Comparison between the results of new method (Bio-computing) and Little's law

| Technique / Average Performance | Bio-computing | Little's Law | Difference |
|---|---|---|---|
| Utilization | 0.695720367 | 0.697068404 | 0.001348037 |
| Expected idle time | 0.304279633 | 0.302931596 | |

The comparison determine the difference between new method (Bio-computing) and Little's law in percentage is **0.134%** for utilization (Busy) and the idle time as shown in Figure 10.

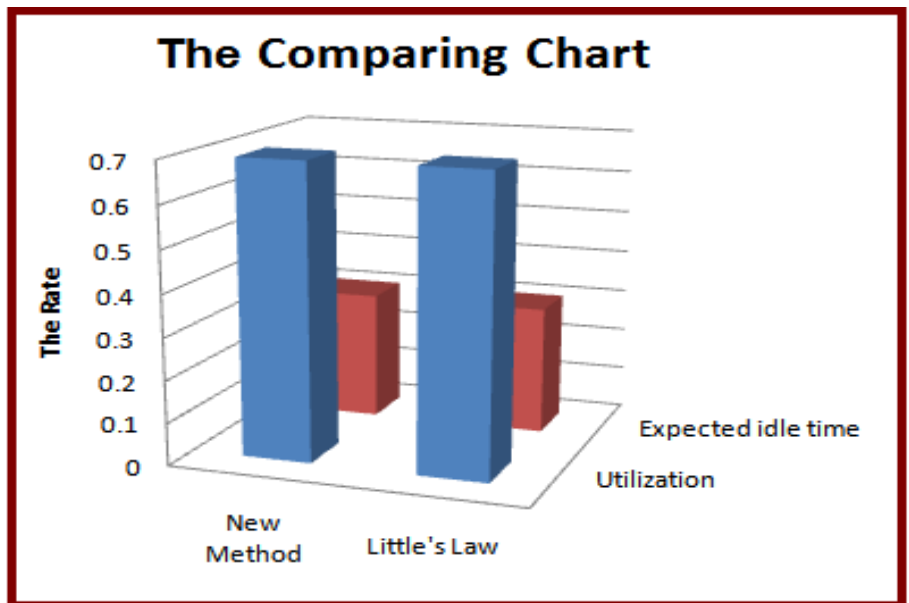

Figure 10: Shows chart for the difference between TCP performance of the new method (Bio-Computing) and Little's law

## 6. CONCLUSIONS

The real-time implementation for BIOKM (server-client model) and its applications to measure the TCP performance using a new method (Bio-computing) give the following conclusions:

**A.** Comparing the results of new method (Bio-computing) with Little's law is very close with difference $\cong 0.13\%$, i.e. dependence the Bio-computing as new method for measuring the real-time TCP performance of IP network.

**B.** Implementing BIOKM application appears obtaining a phylogenetic tree using the NJ method produce a unique tree because the new method base on wisdom results under the principle of minimum evolution.

**C.** Measuring TCP performance using Bio-computing technique especially molecular calculation provides wisdom results and it is possible to exploit all facilities of phylogenetic analysis to obtain better performance and throughput.

**E.** There are many factors affecting on BIOKM and that will reflect on the TCP performance such as:
   **i.** The dynamic parameters of network, e.g. data rate, speed, etc.
   **ii.** Constant factors such as antivirus and memory size which make the system to be slower.
   **iii.** Hub structure (hub only or hub plus firewall), where the firewall doesn't allow KMCS to connect with KMSS.
   **iv.** The Central Processor Unit **CPU** temperature makes the system to be slower.

**Authors**

**Ayad Ghany Ismaeel** is currently an assistant professor of computer science at department of Information Systems Engineering at Erbil Technical College-Iraq. He received his Ph.D. degree in computer science from University of Technology at Baghdad- Iraq in 2006. His research interest's mobile and IP networks, Web application, GPS, GIS techniques, distributed systems and distributed databases. He is lecturer in postgraduate of few universities in MSc and PhD courses in computer science and software engineering from 2007 till now in Kurdistan-Iraq. He is Editorial Board Member of International Journal of Distributed and Parallel Systems (IJDPS), as well as Program Committee Member of conferences related to AIRCC world wide.

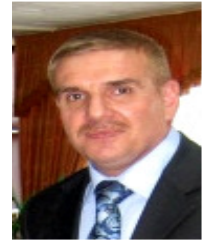

**Suha Adham Abdul-Rahman** is received a B.S. in Computer Science from University of Technology, Iraq in 1991, received an M.S. in Data Base-Cryptography interactive System from University of Technology, Iraq in 1999, and a Ph.D. in BIO-Computing from University of Technology, Iraq in 2004. She has computer science professional/Bio-Computing with more than 15 years of experience in teaching at university level as Over 12 years experience teaching Computer Science at the university of Technology, Baghdad-Iraq. A year experience teaching Computer Science at the Applied Science university, IT Faculty, Amman, Jordan and 4 years experience teaching Computer Science at the Salahddin University, Erbil-Iraq, as well as Five years experience in managing many civic education projects for international universities, organizations and institutions including American University s Center of Global Peace, USAID, DAI, Telecom/Telematique, Inc (T/TI) which is Formed in 1987 to promote development of telecommunications and IT in developing countries.

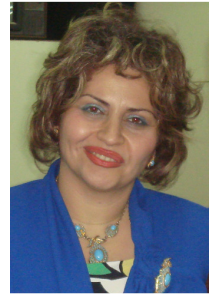